\documentclass[aps,prb,reprint,superscriptaddress,amsmath,amssymb,floatfix]{revtex4-2}
\usepackage{graphicx}
\usepackage{dcolumn}
\usepackage{bm}
\usepackage{xcolor}
\usepackage{soul}
\definecolor{darkred}{rgb}{0.5, 0.0, 0.0}
\definecolor{darkgreen}{rgb}{0.0, 0.5, 0.0}

\usepackage{soul}
\usepackage{placeins}
\usepackage{float}
\usepackage{graphicx}
\usepackage{dcolumn}
\usepackage{bm}

\begin{document}

\title{Exploring the Finite-Temperature Behavior of Rydberg Atom Arrays: A Tensor Network Approach}

\author{Yuzhou Han}
\affiliation{CAS Key Laboratory of Quantum Information, University of Science and Technology of China, Hefei 230026, People's Republic of China}

\author{Hao Zhang}
\affiliation{CAS Key Laboratory of Quantum Information, University of Science and Technology of China, Hefei 230026, People's Republic of China}

\author{Lixin He}
\email{helx@ustc.edu.cn}
\affiliation{CAS Key Laboratory of Quantum Information, University of Science and Technology of China, Hefei 230026, People's Republic of China}
\affiliation{Institute of Artificial Intelligence, Hefei Comprehensive National Science Center, Hefei, 230088, People's Republic of China}
\affiliation{Hefei National Laboratory, University of Science and Technology of China, Hefei 230088, China}

\date{\today}


\begin{abstract}

Rydberg atom arrays have emerged as a powerful platform for experimental research and a challenging subject for theoretical investigation in quantum science. In this study, we investigate the finite-temperature properties of two-dimensional square-lattice Rydberg atom arrays using the projected entangled pair states (PEPS) method. By analyzing the thermal behavior of systems in the checkerboard and striated phases, we extract critical exponents and identify phase transition characteristics. Our results confirm that the checkerboard phase transition belongs to the 2D Ising universality class, while the striated phase exhibits critical exponents that deviate from known universality classes, possibly due to finite-size effects. These findings provide theoretical insights into the thermal stability of quantum phases in Rydberg atom arrays and offer valuable guidance for future experimental efforts.

\end{abstract}
\maketitle


\section{Introduction}

Rydberg atom arrays, composed of cold neutral atoms trapped in optical tweezers \cite{Thompson_2022_Trapping, Anderegg_2019_optical_tweezer, kaufma2021_optical_tweezer}, have emerged as a promising experimental platform for exploring quantum many-body physics \cite{Lukin_51-atom, Lukin_256-atom, Browaeys_controlled_Rydberg_atoms}. When driven by a detuned laser field, the neutral atoms in these arrays are excited to the Rydberg state \cite{Morgado_Interacting_Qubits}, where they experience strong dipole-dipole interactions \cite{Lukin_Fast_Quantum_Gates}. These long-range interactions give rise to the Rydberg blockade effect \cite{Lukin_Fast_Quantum_Gates, Lukin_Dipole_Blockade, Urban_2009_Rydberg_blockade, Ga_tan_2009_blockade}, which prevents the simultaneous excitation of two nearby atoms, thereby enabling precise control over the system. This blockade effect plays a central role in numerous applications, facilitating the programmable realization and high-fidelity manipulation \cite{Levine_2018High-Fidelity_Control, Levine_2019_High-Fidelity_Gate} of Rydberg atom systems. Furthermore, the competition between laser excitation and the Rydberg blockade creates a rich ground-state phase diagram \cite{Lukin_Complex_Density_Wave_1910, Lukin_Bulk_Boundary}, paving the way for new avenues in quantum simulation and the exploration of complex quantum phenomena.

Due to these unique properties, Rydberg atom arrays are increasingly being used as platforms for both quantum information processing \cite{Saffman_2010_Information_Rydberg, Levine_2018High-Fidelity_Control, Levine_2019_High-Fidelity_Gate} and quantum simulation \cite{Lukin_51-atom, Lukin_256-atom, Keesling_2019_Kibble–Zurek, de_L_s_leuc_2019_symmetry-protected_topological, Scholl_2021_Rydberg_Simulation}. Recent experimental advances have successfully demonstrated the creation of novel quantum phases and phase transitions \cite{Lukin_256-atom} in Rydberg lattice systems.

Theoretical investigations have revealed several remarkable phenomena in Rydberg atom systems, including the existence of quantum critical points \cite{Samajdar_2018, Lukin_Complex_Density_Wave_1910, Lukin_Bulk_Boundary}, floating phases \cite{rader_2019_floating, Chepiga_2021_melting}, and topologically ordered spin liquid phases \cite{Samajdar_2021_Rydberg_kagome, Semeghini_2021topological_Rydberg}. While the ground-state properties of two-dimensional square Rydberg systems have been thoroughly examined \cite{Lukin_Complex_Density_Wave_1910, Lukin_256-atom}, their finite-temperature behavior remains less well understood. In experiments, systems are always subject to finite-temperature conditions, which can significantly impact phase stability. Moreover, richer physical phenomena may emerge at finite temperatures. Therefore, understanding the finite-temperature properties of Rydberg atom arrays is essential for guiding future experimental investigations.

In this work, we study the finite-temperature physics of two-dimensional square Rydberg atom arrays. We begin by examining the zero-temperature phase diagram of the Rydberg array using the tensor network state method, specifically the projected entangled pair states (PEPS) scheme \cite{Liu_2017_TNS, Vanderstraeten_2021_TNS, Liu_2021_TNS}. We quantitatively determine the phase boundaries, which are in excellent agreement with both prior experimental \cite{Lukin_256-atom} and numerical \cite{Rourke_2023_Rydberg_TNS} results. Additionally, we uncover a parameter region corresponding to $Z_3$ boundary-ordered phases in the ground-state phase diagram.

Recently, the PEPS scheme has been extended to study the finite-temperature properties of many-particle systems \cite{zhang_2024_finite-T_TNS}. We then analyze the finite-temperature behavior of the Rydberg atom array using this scheme. Our finite-temperature analysis primarily focuses on the parameter regions corresponding to the checkerboard \cite{Lukin_Complex_Density_Wave_1910, Lukin_256-atom} and striated \cite{Lukin_Complex_Density_Wave_1910, Lukin_256-atom} phases. For the systems in the checkerboard ground-state phase, we calculate the temperature-dependent properties and determine the critical exponents using finite-size scaling. Our results suggest that the finite-temperature phase transition in the checkerboard ground-state phase belongs to the 2D Ising universality class \cite{Nightingale_2D_Ising}. Furthermore, we explore how Hamiltonian parameters affect the finite-temperature behavior and phase stability. In the striated ground-state phase, we conduct similar analyses, obtaining the temperature-dependent curves of observables and the thermal transition critical exponents.

The paper is organized as follows: In Sec.~\ref{Model}, we introduce the Rydberg Hamiltonian that we investigate in this work. We briefly describe the numerical method employed in Sec.~\ref{PEPS_Method}. The zero-temperature phase diagram is presented in Sec.~\ref{GROUND-STATE}. We then show our results on finite-temperature behavior and describe the critical phenomena in Sec.~\ref{FINITE-TEMPERATURE}. Finally, we summarize in Sec.~\ref{Summary}.

\section{Methods}

\subsection{Model of the Rydberg systems}
\label{Model}

In Rydberg atom arrays, the atoms are excited by a laser field into Rydberg states, leading to long-range dipole-dipole interactions \cite{Browaeys_controlled_Rydberg_atoms} between the atoms.  The precise form of these interactions is determined by the experimental setup used to create the Rydberg atom array \cite{Browaeys_controlled_Rydberg_atoms, Lukin_256-atom}.
Typically, the system is engineered to exhibit a Van der Waals (VDW) interaction \cite{Lukin_Fast_Quantum_Gates}, which decays proportionally to $1/r^6$, where $r$ is the distance between the atoms.

We consider the following Hamiltonian, which describes a system of Rydberg atoms arranged in a two-dimensional square lattice of size $N \equiv L_x \times L_y$ with open boundary conditions (OBC) \cite{Lukin_Complex_Density_Wave_1910}:
\begin{align}
H_{\text{Ryd}} &= \sum_{i=1}^{N} \frac{1}{2} \left( |g\rangle_i \langle r| + |r\rangle_i \langle g| \right) - \frac{\delta}{\Omega} |r\rangle_i \langle r| \notag \\
&\quad + \frac{1}{2} \sum_{i \neq j} \left( \frac{R_b}{a} \right)^6 \bigg/ \left( \frac{||\mathbf{x}_i - \mathbf{x}_j||}{a} \right)^6 |r\rangle_i \langle r| \otimes |r\rangle_j \langle r|.
\label{eq:hamil}
\end{align}
The units are chosen such that the reduced Planck constant is $\hbar = 1$. The lattice constant is denoted by $a$, and $i$ indexes the lattice sites at positions $\mathbf{x}_i$. Each site has a ground state $|g\rangle_i$ and a Rydberg state $|r\rangle_i$. The system is driven by an external coherent laser field with Rabi frequency $\Omega$ and detuning $\delta$. The Van der Waals (VDW) interactions between atoms in Rydberg states are given by $V(x) = \frac{C_6}{||\mathbf{x}_i - \mathbf{x}_j||^6}$, where $C_6$ is the interaction strength. The Rydberg blockade radius $R_b$ \cite{Lukin_Complex_Density_Wave_1910} is defined by $V(R_b) \equiv \Omega$, which separates the regime where the interaction dominates over the Rabi frequency, preventing neighboring atoms from being simultaneously excited to Rydberg states. The ratio ${R_b}/{a}$ defines the effective blockade range. In experiments, the lattice spacing $a$ is controlled to tune the interactions.

$H_{\text{Ryd}}$ is parameterized by two free parameters: $\delta/\Omega$ and $R_b/a$. The parameter $\delta/\Omega$, which couples to $|r\rangle_i \langle r|$, acts as a longitudinal field in the Hamiltonian. The parameter $R_b/a$, which is coupled to the Rydberg interaction term, governs the Van der Waals (VDW) interactions in the system.
In the regime of small \(\delta/\Omega\), the dominant contribution arises from the Rabi term, leading to a low occupancy of the Rydberg state and an absence of any excitation pattern, resulting in a disordered phase\cite{Lukin_Complex_Density_Wave_1910}. As \(\delta/\Omega\) increases, the system  favors occupation of the Rydberg state, inducing the system to maximize the number of excited atoms. However, the VDW interaction suppresses nearby excitations. As the parameter \({R_b}/{a}\) increases, the distance between Rydberg atoms becomes larger in the excitation pattern. This competition between these two parameters leads several ground-state ordered phases with different spatial symmetries\cite{Lukin_Complex_Density_Wave_1910, Rourke_2023_Rydberg_TNS}.

\subsection{PEPS Method for Ground-State and Finite-Temperature Properties} \label{PEPS_Method}

The model Hamiltonian in Eq.~\ref{eq:hamil} features long-range and frustrated interactions, making it challenging to solve. To investigate the phase diagram of the model, we employ the PEPS method. The ground state wave functions are represented by PEPS on the $N = L_x \times L_y$ square lattices with open boundary conditions (OBC),
\begin{equation}
\left|\Psi\right\rangle = \sum_{s_1 \cdots s_N = 1}^{d} \text{Tr} (A_1^{s_1} A_2^{s_2} \cdots A_N^{s_N}) \left| s_1 s_2 \cdots s_N \right\rangle,
\end{equation}
where the tensor $A_i^{s_i} = A_i(r, l, u, d, s_i)$ is a five-index tensor located at site $i$. $s_i$ is the physical index, and $r$, $l$, $u$, $d$ are the virtual bonds of the PEPS, with a bond dimension $D$. In this study, the ground states are obtained using PEPS with bond dimension $D = 4$, which gives well-converged results.
Because the Hamiltonian involves long-range interactions, the imaginary time evolution with a simple update method \cite{PhysRevLett.101.090603} is not suitable for optimization. Instead, we optimize the wave functions using a stochastic reconfiguration (SR) method \cite{Sorella_1998_SR, Sorella_2001_SR}.

Recently, some of the authors of this paper extended the PEPS  method to finite temperature~\cite{zhang_2024_finite-T_TNS}.
For a quantum system described by a Hamiltonian $H$, the (unnormalized) thermal state at temperature $T= \frac{1}{\beta}$ is given by:
$$
\rho_{\beta} = e^{-\beta H}.
$$
This can be rewritten as:
$$
\rho_\beta = e^{-\frac{\beta}{2} H} I e^{-\frac{\beta}{2} H},
$$
where $I$ is the identity operator, corresponding to the infinite-temperature thermal state (i.e., $\beta = 0 $).

To map the density matrix \( \rho \) into a vector, we apply the vectorization operator\cite{Kshetrimayum_2017_vectoriz,Kshetrimayum_2019_vectoriz}, obtaining:
$$
\rho = \sum_{ss'} \rho_{ss'} |s\rangle \langle s'| \rightarrow \left| \rho \right\rangle_{\sharp} = \sum_{ss'} \rho_{ss'} |s\rangle |s'\rangle.
$$

For the thermal state \( \rho_{\beta} \), we apply the transformation to obtain its vectorized form:
$$
|\rho_\beta\rangle_{\sharp} = e^{-\frac{\beta}{2} H \otimes I} e^{-\frac{\beta}{2} I \otimes H^{T}} |I\rangle_{\sharp} = e^{-\frac{\beta}{2} \mathcal{H}} |I\rangle_{\sharp},
$$
where \( |I\rangle_{\sharp} \) is the vectorized infinite-temperature thermal state, and \( \mathcal{H} = H \otimes I + I \otimes H^{T} \).

We use a projected entangled pair operator (PEPO) \cite{RevModPhys.93.045003} to represent the density matrix. The vectorization of the PEPO combines the two physical indices \( |s_i\rangle \) and \( |s'_i \rangle \) into a single physical index, \( |S_{i}\rangle = |s_i\rangle |s'_i \rangle \), of dimension \( d^2 \). After vectorization, the density matrix is expressed as a PEPS:
\begin{equation}
\left|\rho\right\rangle_{\sharp} = \sum_{S_{1}, \cdots, S_{N}=0}^{d^2-1} \operatorname{Tr}\left(T_{1}^{S_{1}} \cdots T_{N}^{S_{N}}\right) \left|S_{1} \cdots S_{N}\right\rangle.
\end{equation}
We then apply the SR method \cite{Sorella_1998_SR,Sorella_2001_SR}  to evolve from the infinite temperature state $|I\rangle_{\sharp}$ to obtain the thermal state at temperature $T = 1/\beta$.
For the finite-temperature calculations, the bond dimension $D$=7 is used.

\section{Results and discussion}

\subsection{Ground-State Phase Diagram}
\label{GROUND-STATE}

\begin{figure}[!tb]
    \centering
    \includegraphics[width=0.9\linewidth]{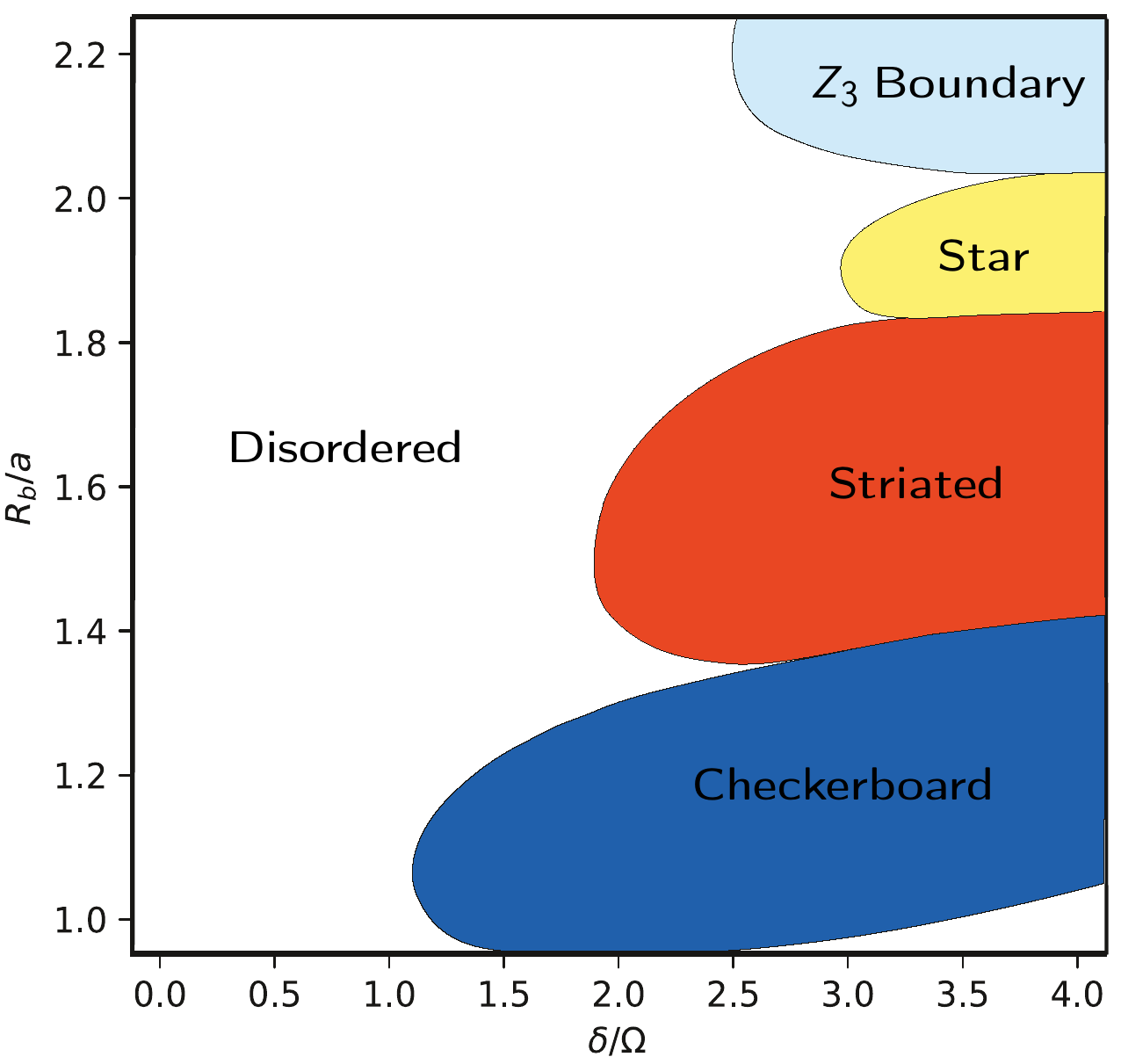}
    \caption{Ground-state phase diagram of the 2D Rydberg Hamiltonian with OBC.  }
    \label{fig:Phase Diagram}
\end{figure}

We first examine the ground state of the Rydberg Hamiltonian. To obtain the phase diagram, we scanned the parameter ranges of $R_b/a$ and $\delta/\Omega$. The PEPS scheme \cite{Liu_2017_TNS, Vanderstraeten_2021_TNS, Liu_2021_TNS} is employed to obtain the ground state on a $15 \times 15$ lattice. In order to distinguish between the different phases and explore their spatially ordered properties, we calculate the symmetrized Fourier transform of the Rydberg excitation density in momentum space, which is defined as
\begin{equation}
\langle n(\mathbf{k}) \rangle = \widetilde{\mathcal{F}}(k_x, k_y) \equiv \frac{1}{2} \left( \mathcal{F}(k_x, k_y) + \mathcal{F}(k_y, k_x) \right)
\end{equation}
where
\begin{equation}
\mathcal{F}(k_x, k_y) = \frac{1}{N} \sum_i \langle n_i \rangle e^{-i \mathbf{k} \cdot \mathbf{r_i}},
\end{equation}
with $(k_x, k_y)$ in momentum space, $\langle n_i \rangle$ being the density of Rydberg excitations at site $i$, and $N$ being the total number of atoms in the system.

In the considered parameter range, we uncover four bulk phases: disordered, checkerboard, striated, and star. We show the Rydberg excitation densities in real space and their Fourier transforms in momentum space for several representative parameter points in different phases in Fig.~\ref{fig:Real space} in Appendix~\ref{sec:Detail_Ground-state}. The disordered phase does not break any symmetries. The checkerboard phase breaks $Z_2$ translational symmetry, with the order parameter corresponding to $\widetilde{\mathcal{F}}(\pi, \pi)$ \cite{Lukin_256-atom, Lukin_Bulk_Boundary}. The striated phase breaks the $Z_2 \times Z_2$ translational symmetry, and the order parameter is $\widetilde{\mathcal{F}}(\pi, 0)$ \cite{Lukin_256-atom, Lukin_Bulk_Boundary}. The Star phase breaks both $Z_2$ symmetry and $C_4$ rotational symmetry, with $\widetilde{\mathcal{F}}(\pi/2, \pi)$ serving as the corresponding order parameter \cite{Lukin_256-atom, Lukin_Bulk_Boundary}.

The phase diagram of the Rydberg system is schematically shown in Fig.~\ref{fig:Phase Diagram}. The order parameters for different phases are illustrated in Fig.~\ref{fig:Phase Diagram Original} in Appendix~\ref{sec:Detail_Ground-state}. The ground-state phase diagram of 2D square Rydberg atom arrays has been extensively discussed in the literature \cite{Lukin_Complex_Density_Wave_1910, Lukin_256-atom, Rourke_2023_Rydberg_TNS, Lukin_Bulk_Boundary}, and our results for the bulk phases under OBC are consistent with previous numerical \cite{Rourke_2023_Rydberg_TNS} and experimental \cite{Lukin_256-atom} studies.

Besides these four bulk phases, we also identify a $Z_3$ boundary-ordered phase, which is disordered in the bulk but exhibits ordered characteristics at the boundary \cite{Lukin_Bulk_Boundary}, where a Rydberg state excitation occurs every three sites. We choose representative parameter points $\delta/\Omega = 3.5$, $R_b/a = 2.2$ to show the Rydberg excitation density in real space and in momentum space for the phase in Fig.~\ref{fig:Z_3}. The $Z_3$ boundary-ordered phase can be identified using $\widetilde{\mathcal{F}}\left( \frac{\pi}{3}, \frac{\pi}{3} \right)$ as an order parameter.

\subsection{Finite-Temperature Behavior and Thermal Phase Transitions}\label{FINITE-TEMPERATURE}

We investigate the finite-temperature physics of two-dimensional square-lattice Rydberg atom arrays by selecting representative parameter points from both the checkerboard and striated ground-state phases. Using the recently developed extended PEPS algorithm for finite-temperature simulations \cite{zhang_2024_finite-T_TNS}, we numerically determine the thermal equilibrium properties of the Rydberg system. This advanced methodology allows for a comprehensive characterization of the system's finite-temperature properties, covering the entire temperature range from infinite temperature to zero temperature.

\subsubsection{Checkerboard Phase}

The checkerboard phase \cite{Lukin_Complex_Density_Wave_1910, Lukin_256-atom} analogizes to an antiferromagnetic phase, spatially ordered with a twofold degenerate ground-state. At high temperatures, the system behaves as a ``paramagnet'' with featureless spatial excitation. To identify the phase transition, we use the staggered magnetization \cite{Lukin_Complex_Density_Wave_1910, Liquid_Helium} \( m_s \) as the order parameter to detect \(Z_2\)-symmetry breaking, which is defined as:
\begin{equation}
m_s = \langle |M_s| \rangle, \quad
M_s = \frac{1}{N} \sum_i (-1)^{i} \left( n_i - \frac{1}{2} \right),
\end{equation}

\begin{figure}[bp]
    \centering
    \includegraphics[width=0.9\linewidth]{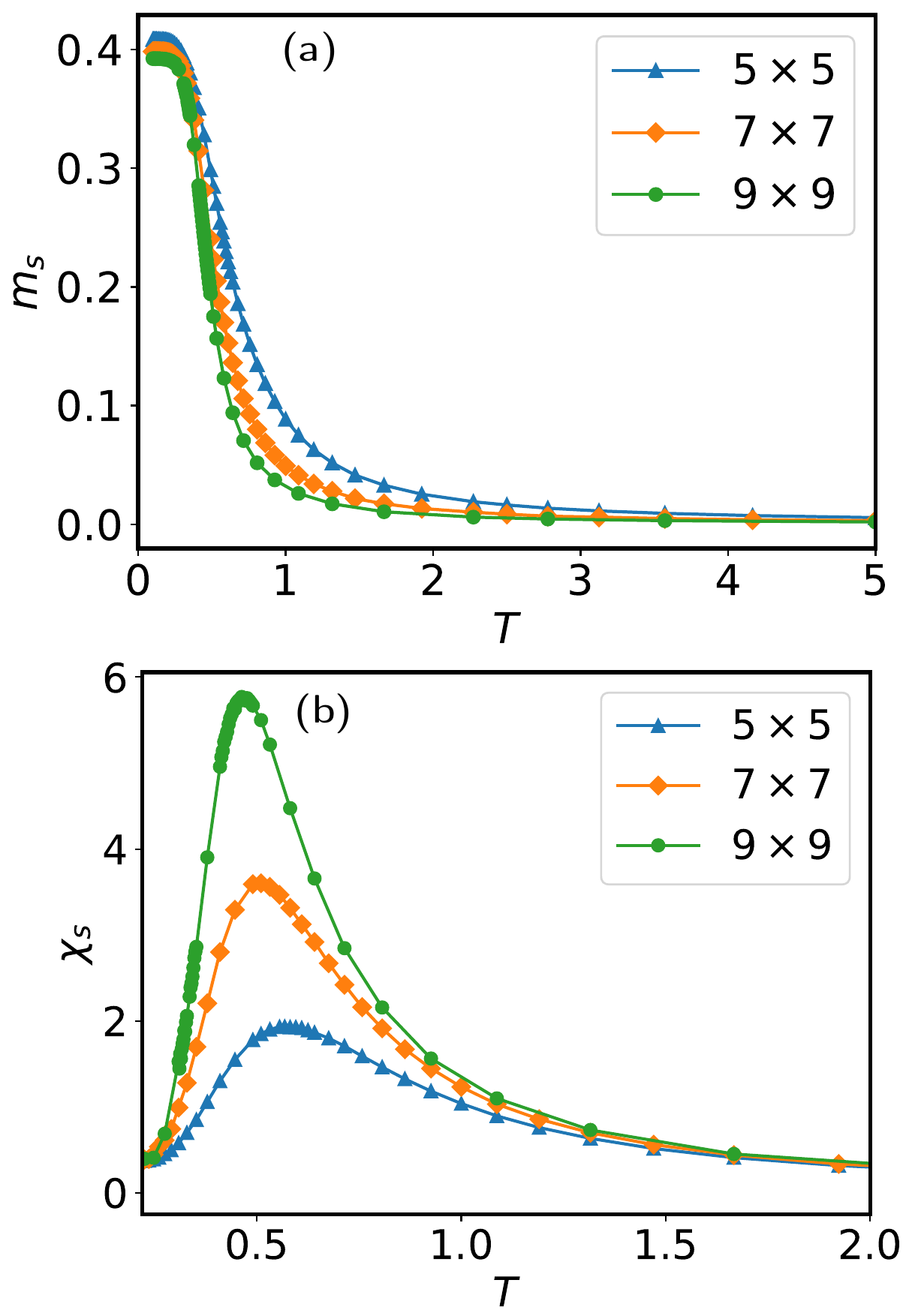}
    \caption{Finite-temperature behavior for (a) staggered magnetization $m_s$ and (b) staggered magnetic susceptibility $\chi_s$ at $\delta/\Omega = 2.0$, $R_b/a = 1.2$ for several system sizes.}
    \label{fig:observable_Checkerboard}
\end{figure}

Subsequently, we introduce the staggered magnetic susceptibility \cite{Finite-T_long-range_Ising} \( \chi_s \), defined as \( \chi_s = \left( \frac{\partial \langle m_s \rangle}{\partial h} \right)_T \). In the Rydberg atom system, the Hamiltonian does not commute with the chosen order parameter \( m_s \), i.e., \( [H_{\text{Ryd}}, m_s] \neq 0 \). In this case, the magnetic susceptibility should be given by the Kubo formula \cite{Humeniuk_2020, Sticlet_2022}. However, this is computationally intractable with our tensor network algorithm.

To simplify the computation, we examine the critical behavior of the magnetic susceptibility using the {\it classical} susceptibility form \cite{Humeniuk_2020, Finite-T_long-range_Ising} \( \chi_{\text{cl}} \), which is given by:
\begin{equation}
\chi_{\text{cl}} = \frac{N}{k_B T} \left( \langle {M_s}^2 \rangle - \langle M_s \rangle^2 \right),
\end{equation}
Although $\chi_{\text{cl}}$ may deviate from $\chi_s$ at extremely low temperatures, it has been shown to provide a good approximation near the phase transition temperature, accurately reflecting the phase transition behavior of the system \cite{Humeniuk_2020, Finite-T_long-range_Ising}. Therefore, in this work, we use $ \chi_{\text{cl}} $ to approximate $ \chi_s $ for analyzing the phase transition.

To investigate the finite-temperature behavior of the system in the checkerboard phase, we choose a representative parameter point $\delta/\Omega = 2.0$, $R_b/a = 1.2$ within the corresponding phase region.
We first benchmark our algorithm against the exact diagonalization (ED) method on a $3 \times 3$ square lattice. As shown in Fig.\ref{fig:Benchmark} in Appendix~\ref{sec:Benchmark}, the results from PEPS are in excellent agreement with those from ED. We then perform simulations on $5 \times 5$, $7 \times 7$, and $9 \times 9$ lattices. The results for the staggered magnetization $ m_s $ and susceptibility $ \chi_s $ are presented in Fig.~\ref{fig:observable_Checkerboard}(a) and Fig.~\ref{fig:observable_Checkerboard}(b), respectively.
The Rydberg excitations in real space at different temperatures on the  $9 \times 9$ lattice can be found in
Fig.~\ref{fig:Cooling}(a).

At high temperatures, the order parameter $m_s$ remains nearly zero. As the temperature decreases, the system transitions into an ordered state, with a rapid growth of $m_s$, accompanied by a peak in $\chi_{\text{cl}}$. For the chosen parameter point, the blockade radius $R_b$ is comparable to the lattice spacing, leading to significant suppression of double occupancy on neighboring sites during the cooling process. As the temperature continues to decrease, the system develops a checkerboard pattern of excited atoms and approaches the ground-state configuration obtained in Sec.~\ref{GROUND-STATE}.

\begin{figure}[tbp]
    \centering
    \includegraphics[width=0.4\textwidth]{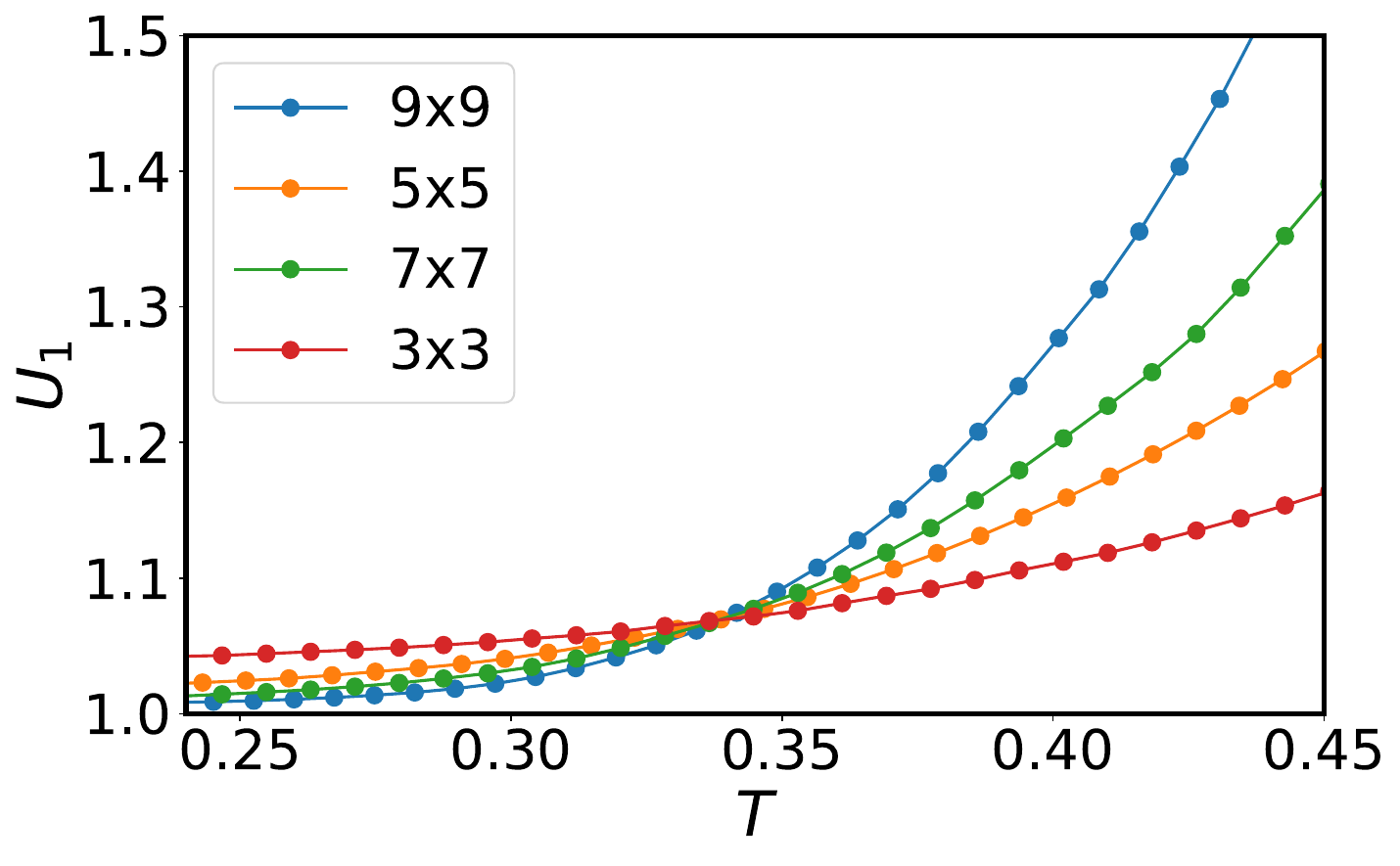}
    \caption{First-order Binder ratio $U_1$ for different system sizes at finite temperature.}
    \label{fig:Checkerboard_BinderRatio}
\end{figure}

To estimate the phase transition temperature in the thermodynamic limit, we use the first-order Binder ratio\cite{Vink_2010_Binder} \(U_1\), which is size-independent at the critical point, defined as:
\begin{equation}
U_1 = \frac{\langle m^{2} \rangle}{\langle |m| \rangle^2}.
\end{equation}
As shown in Fig.~\ref{fig:Checkerboard_BinderRatio}, the Binder ratio \(U_1\) for lattices of different sizes intersects at the same temperature, from which we can extrapolate the critical temperature in the thermodynamic limit to be \(T_c(\infty) = 0.340\).

\begin{figure}[t]
    \centering
    \includegraphics[width=0.9\linewidth]{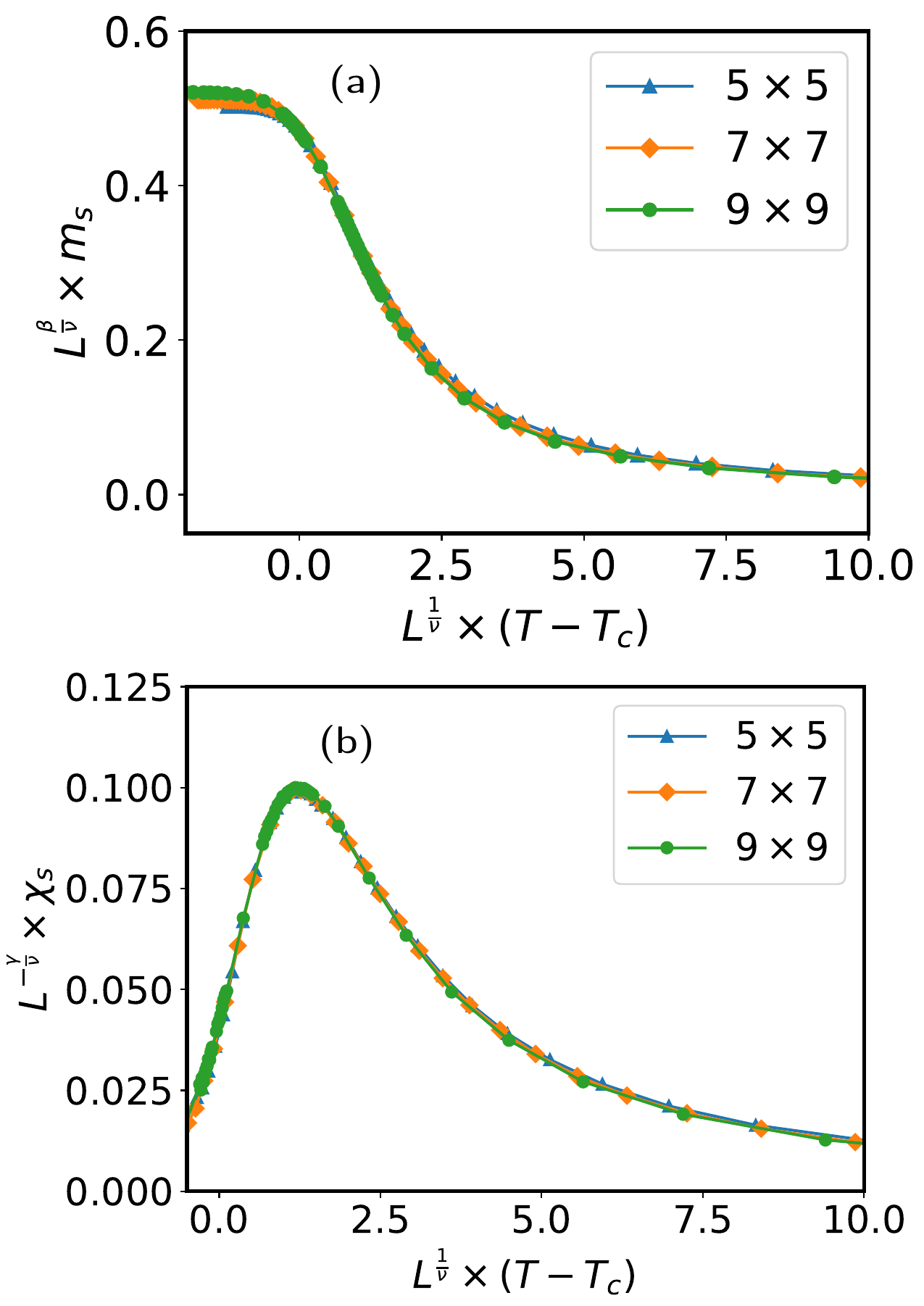}
    \caption{Finite-size scaling data collapse for the checkerboard phase: (a) staggered magnetization $m_s$ and (b) susceptibility $\chi_s$ at $\delta/\Omega = 2.0$ and $R_b/a = 1.2$.
    }
    \label{fig:rescale_Checkerboard}
\end{figure}

According to the finite-size scaling (FSS) theory, the scaling of the magnetization \(m_s\) and susceptibility \(\chi_s\) at different system sizes is given by\cite{Finite-T_long-range_Ising}:
\begin{equation}
m_s = L^{-\frac{\beta}{\nu}} \mathcal{F}_{m_s}\left(L^{\frac{1}{\nu}} (T - T_c)\right),
\end{equation}
\begin{equation}
\chi_s = L^{\frac{\gamma}{\nu}} \mathcal{F}_{\chi}\left(L^{\frac{1}{\nu}} (T - T_c)\right).
\end{equation}
Both $m_s$ and $\chi_{\text{cl}}$ can be fitted very well by $\nu = 0.970$, $\beta = 0.125$, and $\gamma = 1.790$, as illustrated in Fig.~\ref{fig:rescale_Checkerboard}(b)(c) for all three lattice sizes.
The fitted critical exponents closely match those of the two-dimensional Ising universality class \cite{Nightingale_2D_Ising}, with $\nu = 1$, $\beta = \frac{1}{8}$, and $\gamma = \frac{7}{4}$. This indicates that the system, which exhibits a checkerboard phase as its ground state, undergoes a second-order phase transition at finite temperature. Moreover, the transition is in agreement with the 2D Ising universality class.

\begin{figure}[tbp]
    \centering
    \includegraphics[width=0.8\linewidth]{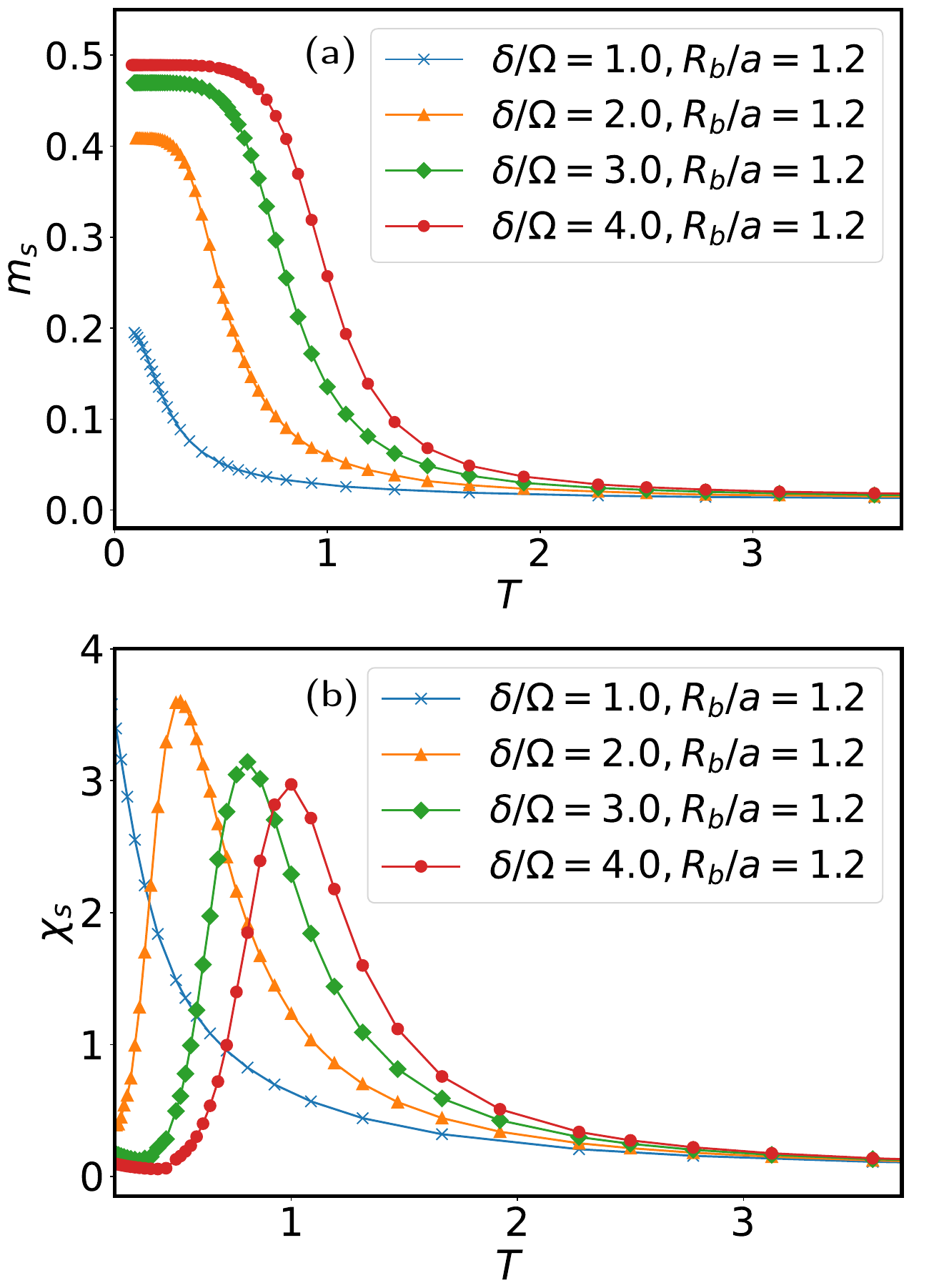}
    \caption{ Finite-temperature behavior of (a) magnetization $m_s$ and (b) staggered magnetic susceptibility $\chi_s$ at $\delta/\Omega = 1.0$, $2.0$, $3.0$, and $4.0$ along $R_b/a = 1.2$, calculated for a $7 \times 7$ lattice.
}
    \label{fig:Contrast_Checkerboard}
\end{figure}


 To further investigate the impact of parameter changes on the finite-temperature behavior, we select four representative points along  $R_b/a = 1.2$, specifically at $\delta/\Omega = 1.0$, $\delta/\Omega = 2.0$, $\delta/\Omega = 3.0$, and $\delta/\Omega = 4.0$. Among these parameters, $\delta/\Omega = 1.0$ is in the disordered phase, whereas the other three points are in the checkerboard phase. The finite-temperature behavior is examined for a system of size $7 \times 7$, as shown in Fig.~\ref{fig:Contrast_Checkerboard}.
 The results indicate that, for $\delta/\Omega = 1.0$, no phase transition occurs. For other parameter values, as $\delta/\Omega$ increases, the critical temperature $T_c$ of the phase transition increases, and the magnetization strength $m_s$ at low temperatures also rises. This behavior is consistent with the variation in the order parameter observed in the ground-state calculations.

\subsubsection{Striated Phase}

\begin{figure}[htbp]
    \centering
    \includegraphics[width=1\linewidth]{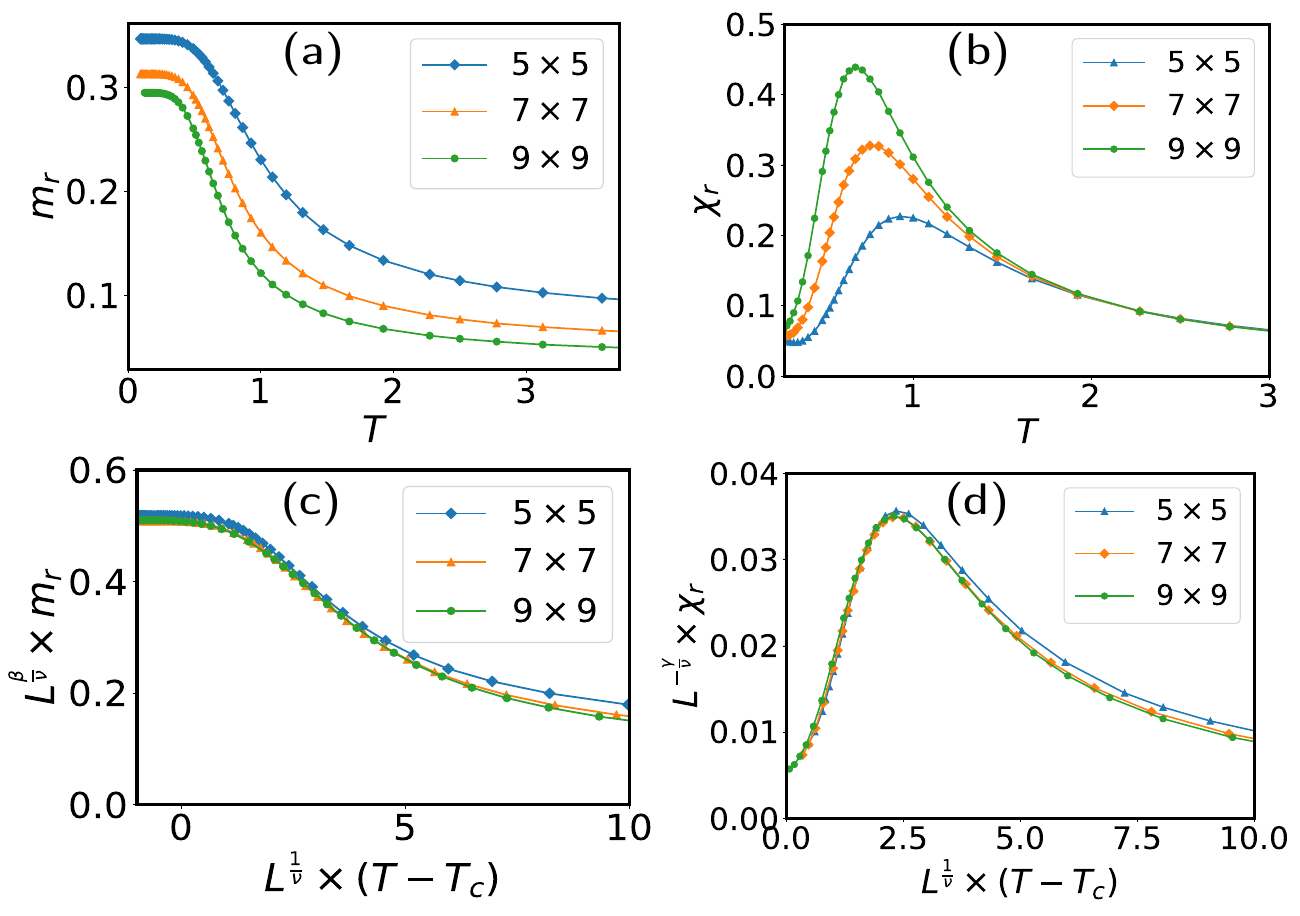}
    \caption{Finite-temperature behavior in the striated ground-state phase: (a) Row magnetization $m_r$; (b) Row magnetic susceptibility $\chi_r$; (c) Data collapse of row magnetization; (d) Data collapse of row susceptibility. The data are calculated for several system sizes at $\delta/\Omega = 3.5$, $R_b/a = 1.6$.
    }
    \label{fig:Striated}
\end{figure}

The Striated phase \cite{Lukin_Complex_Density_Wave_1910,Lukin_256-atom} exhibits a non-zero row magnetization in the ground state, a feature that cannot be derived from the classical Rydberg Hamiltonian \cite{Rourke_2023_Rydberg_TNS}. To describe the thermal phase transition behavior of the system, we use the row magnetization \cite{Lukin_Complex_Density_Wave_1910} $m_r$ as the order parameter, which is defined as:
\begin{equation}
m_r = \langle |M_r| \rangle, \quad
M_r = \frac{1}{N} \sum_i (-1)^{\text{row}(i)} n_i,
\end{equation}
where $\text{row}(i)$ denotes the row index of each Rydberg atom. The order parameter $m_r$ reflects the breaking of the $Z_2 \times Z_2$ symmetry during the cooling process of the Rydberg system. Consequently, we use the following classically approximated formula to compute the corresponding row magnetic susceptibility $\chi_r$:
\begin{equation}
\chi_r = \frac{N}{k_B T} \left( \langle {M_r}^2 \rangle - \langle M_r \rangle^2 \right).
\end{equation}

The representative parameter point is selected as $\delta/\Omega = 3.5$ and $R_b/a = 1.6$. We present the Rydberg excitations in real space at different temperatures in Fig.~\ref{fig:Cooling}(b).
As shown in the figure, the system initially exhibits a thermally disordered state. As the temperature decreases, the system gradually transitions towards an ordered state. For the chosen parameter point, the blockade radius $R_b$=1.6$a$, which not only induces blockade effects on neighboring sites but also significantly suppresses double occupancy on next-nearest-neighbor sites. As the temperature decreases further, the system clearly develops a ``striated'' pattern of excited atoms. At extremely low temperatures, the system approaches the ground-state configuration obtained in Sec.~\ref{GROUND-STATE}.

The corresponding values of $m_r$ and $\chi_r$ as functions of temperature are shown in Fig.~\ref{fig:Striated}(a) and Fig.~\ref{fig:Striated}(b), respectively. We further perform Binder ratio and FSS analyses, as described in the previous section, to examine the critical behavior near the phase transition temperature and extract the critical exponents. The results of these analyses are presented in Fig.~\ref{fig:Striated}(c) and Fig.~\ref{fig:Striated}(d).
By tuning the parameters to $\beta/\nu = 0.280$ and $\gamma/\nu = 1.200$ at $T = 0.300$, the data for different lattice sizes collapse onto nearly the same curves, as illustrated in Fig.~\ref{fig:Striated}(c) and Fig.~\ref{fig:Striated}(d).

Compared to the checkerboard phase, the results at higher temperatures exhibit slight deviations. One possible reason is that, in models exhibiting $\mathbb{Z}_2 \times \mathbb{Z}_2$ symmetry breaking, smaller systems are more strongly affected by boundary conditions and finite-size effects. Consequently, critical exponent estimates obtained from FSS in such systems may be subject to systematic inaccuracies. Additionally, the extracted critical exponents, $\beta/\nu$ and $\gamma/\nu$, do not match those of any well-established universality class. This discrepancy may result from the limited system sizes accessible in our simulations, leading to effective exponents that deviate from their true asymptotic values. Future studies involving larger system sizes could provide more accurate insights into the nature of these ground-state phases.

\section{Summary}
\label{Summary}

We investigate the ground-state and finite-temperature properties of two-dimensional square-lattice Rydberg atom arrays using the extended PEPS methods.
We first revisit the zero-temperature phase diagram, determining phase boundaries and identifying a parameter regime exhibiting $Z_3$ boundary-ordered phases. We then analyze the finite-temperature properties of systems in the checkerboard and striated phases. For the checkerboard phase, we compute observables and extract critical exponents via finite-size scaling analysis, confirming that the phase transition belongs to the two-dimensional Ising universality class.
In the striated phase, we perform a similar analysis. However, the extracted critical exponents do not match those of well-established universality classes, which may be attributed to strong finite-size effects.
Our results demonstrate the effectiveness of tensor network methods in studying finite-temperature properties of Rydberg atom arrays and provide theoretical benchmarks for future experimental investigations.

\begin{acknowledgments}
This work was supported by the National Natural Science Foundation of China under Grant Nos. 12134012, the Strategic Priority Research Program of the Chinese Academy of Sciences under Grant No. XDB0500201, as well as the Innovation Program for Quantum Science and Technology under Grant No. 2021ZD0301200. The numerical calculations in this study were performed on the ORISE Supercomputer and the USTC HPC facilities.
\end{acknowledgments}

\appendix

\section{Further Details on Ground-State Results}
\label{sec:Detail_Ground-state}

\begin{figure*}[!htbp]
    \centering
    \includegraphics[width=0.9\textwidth]{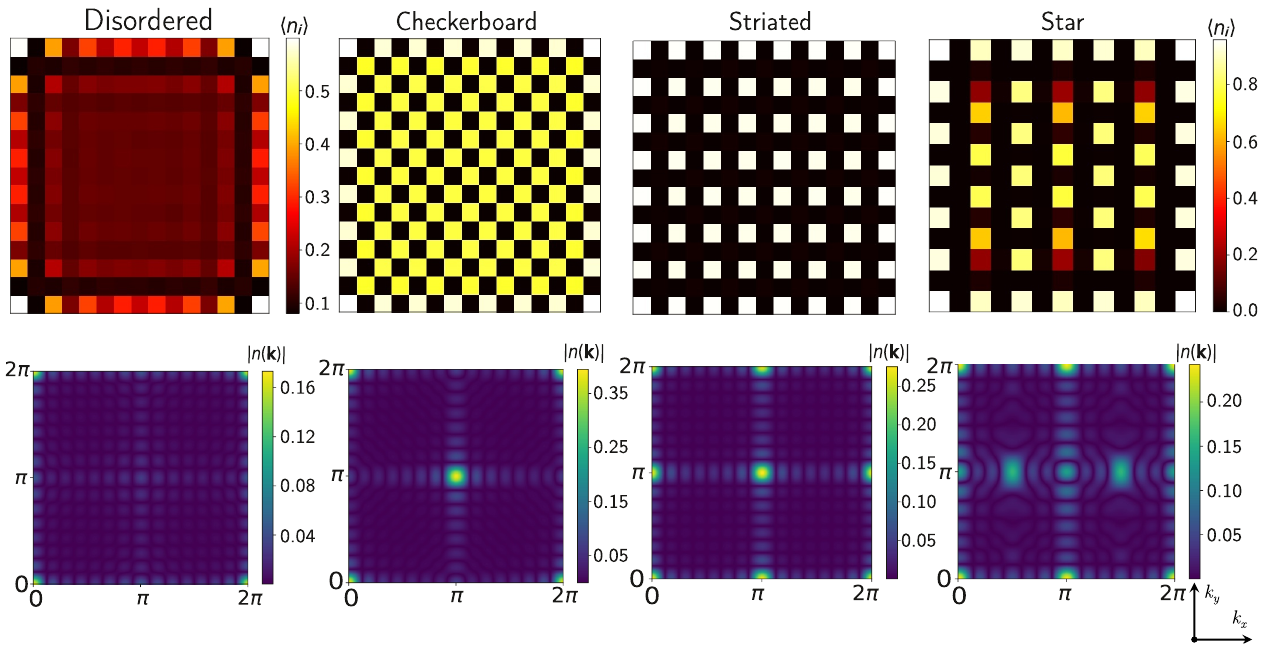}
    \caption{Rydberg excitation densities (top) of bulk phases and their Fourier transforms in momentum space (bottom). The four phases shown are the disordered, checkerboard, striated, and star phases.
     }
    \label{fig:Real space}
\end{figure*}

\begin{figure}[tb]
    \centering
    \includegraphics[width=\linewidth]{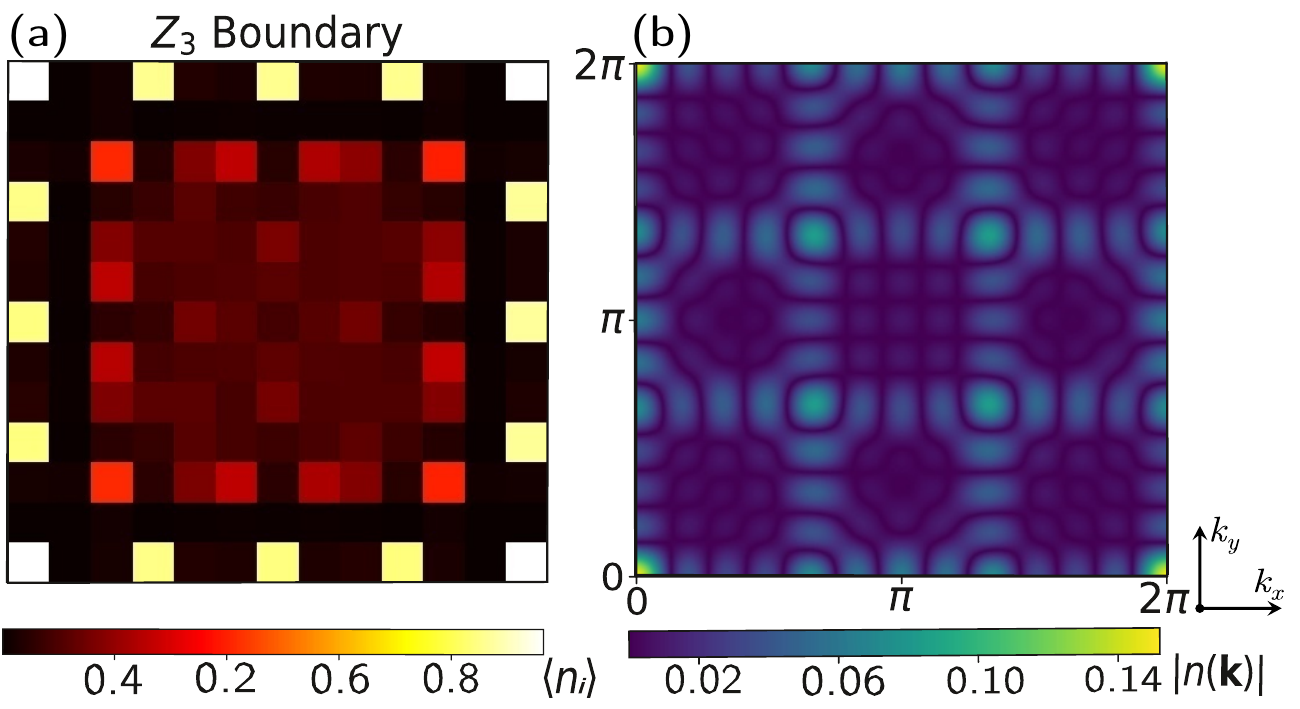}
    \caption{Rydberg excitations of the $Z_3$ boundary-ordered phase and its Fourier transform in momentum space.
 }
    \label{fig:Z_3}
\end{figure}

In the ground-state study, we uncover four bulk phases: disordered, checkerboard, striated, and star. In Fig.~\ref{fig:Real space}, we show the real-space distributions of the Rydberg excitation $\langle n_i \rangle$ and the Fourier transform $\langle n(\mathbf{k}) \rangle$ in momentum space. We select the parameter point $\delta/\Omega = 1.0$, $R_b/a = 1.2$ to illustrate the disordered phase, and use $\delta/\Omega = 2.0$, $R_b/a = 1.2$ for the checkerboard phase. The striated phase is represented by $\delta/\Omega = 3.5$, $R_b/a = 1.6$, and the Star phase is shown by $\delta/\Omega = 3.75$, $R_b/a = 1.9$. The characteristics of these phases in real space and momentum space are consistent with previous literature \cite{Lukin_256-atom, Rourke_2023_Rydberg_TNS}.
Additionally, we identify a $Z_3$ boundary-ordered phase, which is disordered in the bulk but exhibits order at the boundary, as shown in Fig.~\ref{fig:Z_3}(a). The Fourier transform of the Rydberg excitation exhibits peaks at $(\frac{\pi}{3}, \frac{\pi}{3})$.

\begin{figure}[tb]
    \centering
    \includegraphics[width=1\linewidth]{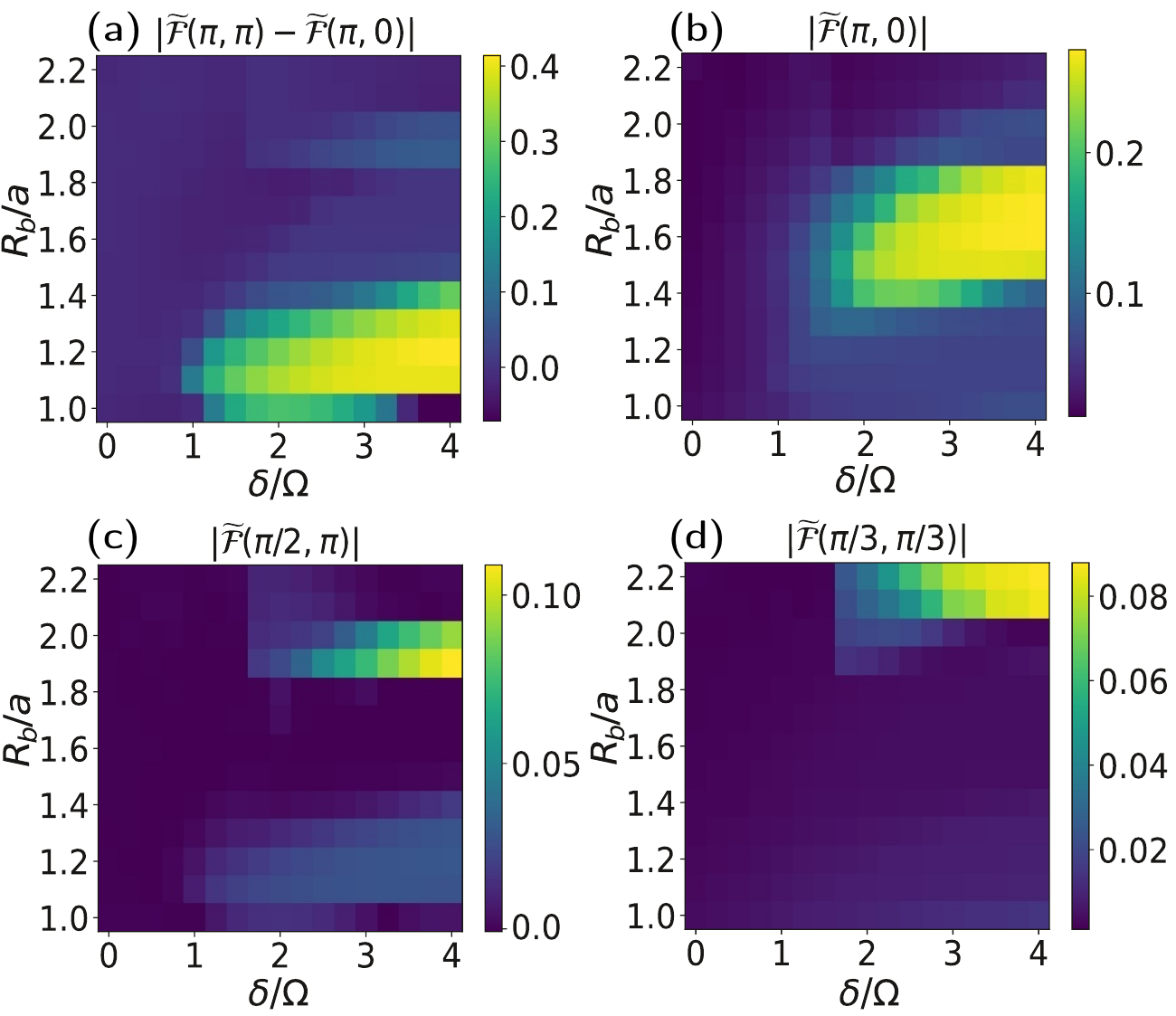}
    \caption{The calculated order parameters for (a) checkerboard phase, (b) striated phase, (c) star phase, and (d) the $Z_3$ boundary-ordered phase.
     }
    \label{fig:Phase Diagram Original}
\end{figure}

We use the following order parameters to distinguish the phases: $\widetilde{\mathcal{F}}(\pi, \pi) - \widetilde{\mathcal{F}}(\pi, 0)$ to identify the checkerboard phase, $\widetilde{\mathcal{F}}(\pi,0)$ for the striated phase, $\widetilde{\mathcal{F}}(\pi/2,\pi)$ for the Star phase and $\widetilde{\mathcal{F}}(\frac{\pi}{3}, \frac{\pi}{3})$ for the $Z_3$ boundary-ordered phase. The calculated order parameters for the $(R_b/a, \delta/\Omega)$ parameters are shown in Fig.~\ref{fig:Phase Diagram Original}. Using these order parameters, we obtain the phase diagram of the Rydberg system, which is schematically shown in Fig.~\ref{fig:Phase Diagram}.


\section{Benchmark of Finite-temperature PEPS Algorithm}\label{sec:Benchmark}

\begin{figure}[tb]
    \centering
    \includegraphics[width=\linewidth]{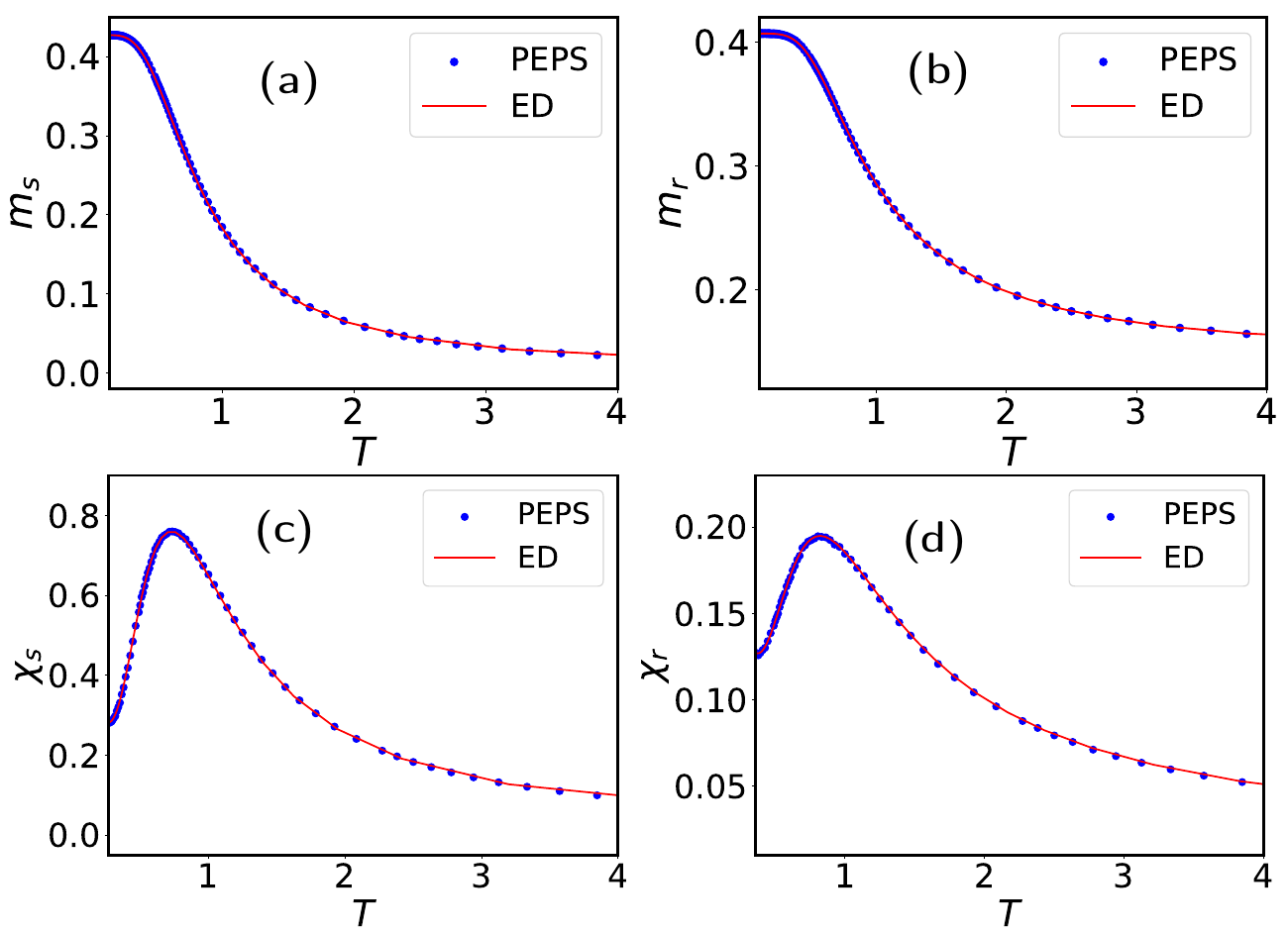}
    \caption{Compare the PEPS algorithm with the ED method on a $3 \times 3$ lattice. Figures (a) and (c) compare the staggered magnetization and staggered magnetic susceptibility in the checkerboard phase, respectively. Figures (b) and (d) compare the row magnetization and row magnetic susceptibility in the striated phase, respectively.}
    \label{fig:Benchmark}
\end{figure}

To validate the accuracy of our finite-temperature PEPS algorithm for the Rydberg system, we compared the results with the exact values obtained from ED in a $3 \times 3$ lattice. We selected a parameter point in the checkerboard phase, specifically $\delta/\Omega = 2.0$, $R_b/a = 1.2$, and a parameter point in the striated phase, namely $\delta/\Omega = 2.5$, $R_b/a = 1.4$, then calculated the magnetization and susceptibility of these two systems in finite temperature. The results obtained from our algorithm are in excellent agreement with those from ED, as shown in Fig.~\ref{fig:Benchmark}. This benchmark demonstrates the accuracy and reliability of our approach.

\section{Rydberg Excitations at Varying Temperatures}\label{sec:Cooling}

\begin{figure*}[htbp]
    \centering
    \includegraphics[width=\linewidth]{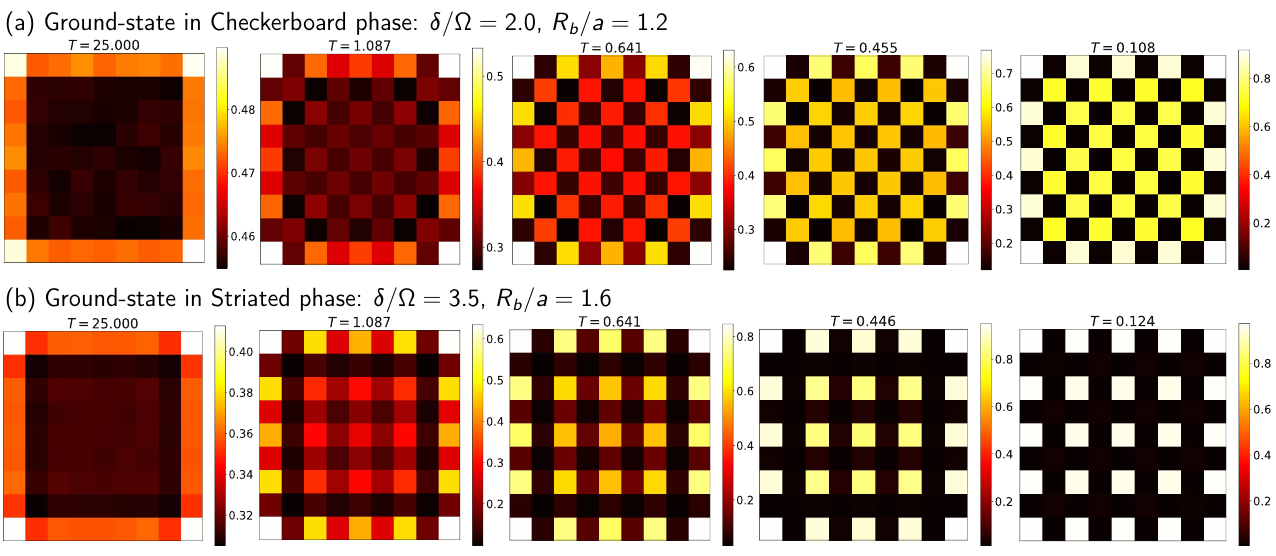}
    \caption{Rydberg excitation at different temperatures: (a) in the checkerboard ground-state phase and (b) in the striated ground-state phase.
}
    \label{fig:Cooling}
\end{figure*}

Figures~\ref{fig:Cooling}(a) and (b) depict the Rydberg excitations in real space for the checkerboard and striated ground-state phases at different temperatures, respectively. As shown in the figures, the system initially exhibits a thermally disordered state. As the temperature decreases, the system gradually transitions toward the ordered states.

%

\end{document}